\documentclass{article}
\usepackage{fullpage}
\usepackage{graphicx}
\usepackage{authblk}

\let\oldthebibliography\thebibliography
\renewcommand\thebibliography[1]{
  \oldthebibliography{#1}
  \setlength{\parskip}{0pt}
  \setlength{\itemsep}{0pt plus 0.3ex}
}

\begin{document}

\title{Sigil3D: A Crowdsourcing Platform for Interactive 3D Content}
\author[1]{Andrea Barillari}
\author[1]{Daniele Bernardini\thanks{daniele.bernardini@intranetstandard.com }}
\author[2]{Pierluigi Crescenzi\thanks{pierluigi.crescenzi@unifi.it}}
\affil[1]{Intranet Standard GmbH, Ottostrasse 3, 80333 Munich, Germany}
\affil[2]{University of Florence,  50134 Florence, Italy}

\date{}
\maketitle

\begin{abstract}
In this paper we propose applying the crowdsourcing approach to a software platform that uses a modern and state-of-the-art 3D game engine. This platform could facilitate the generation and manipulation of interactive 3D environments by a community of users producing different content such as cultural heritage, scientifice virtual labs, games, novel art forms and virtual museums. 
\end{abstract}

\section{Introduction}

Crowdsourcing is known as the process of designing and developing a web-based software platform and obtaining content for it through the contributions of a community, including people who are not part of the entity that conceived the platform itself. One of the most popular crowdsourcing platforms is Wikipedia, an online encyclopaedia, which proved during the years that the crowdsourcing model works efficiently without any significant loss of quality (indeed the average article quality increases as it goes through iterations of edits~\cite{JavLop2010}). Another well-known example of a very successful crowdsourcing platform is YouTube, which currently allows billions of users to share originally-created videos~\cite{BurGre2013}.

On the other hand, in the last twenty years we have seen an impressive development of 3D game engines, starting from the Doom Engine, produced in the early nineties, which really ushered in a new era of game design and play, arriving to the third version of Unreal Engine, presented in 2004, which was a groundbreaking engine capable of handling high quality graphics and powerful user interface while maintaining good usability~\cite{LewJac2002}. All these game engines, however, were quite low-level tools and most of them were proprietary software, meaning that a ``normal'' user could not easily master them in order to produce new content. This drastically changed in the last couple of years with the introduction of the latest modern state-of-the-art game engines, such Unity 5 and Unreal Engine 4 ~\cite{unity}, \cite{unreal}. These kinds of engines are currently used by millions of game developers (although they also support functionalities that can be used for non-gaming purposes).

The goal of this paper is to apply the crowdsourcing approach to a software platform that uses a modern 3D game engine, in order to generate interactive 3D environments by a community of users. This platform could be used to produce content in different fields, from cultural heritage, virtual museums and art galleries to science classes and labs. 

This platform should easily host virtual reconstructions of archaeological sites and  monuments, historical buildings and, in general, works of art. These reconstructions should not only be a virtual representation of the current status, but also a representation of their original status. The platform should work with virtual environments that will be completely explorable in first and third person (for instance, using avatars). The environments should also be populated with representations of the people living in a specific epoch. In the hands of expert historians, archaeologists and architects it should be possible to recreate high fidelity 3D reconstructions, which can be shared among academics to improve research work and course quality. With significant crowdsourcing support, different environments should be joined and theoretically the whole world (or at least the most ``important'' parts of it) could be virtualised and reconstructed. In other words, this platform should become a huge 3D cultural heritage encyclopaedia and repository. 

This platform could also be used to enhance the learning and teaching process by creating low-cost alternative solutions of real labs to educators. Through this platform one can design and create 3D virtual labs to simulate scientific experiments and understand the scientific phenomena and the principles underneath \cite{Vlabs}.

Finally, the platform could be used for gaming purpose due the game engines intrinsic nature.
As a first step towards this ambitious goal we have produced a standalone explicit crowdsourcing system, called \textit{Sigil3D}, which can be considered a \textit{proof of concept} or a prototype implementation with the purpose of verifying the potentials of the platform. As a preliminary validation of this proof of concept, we will also shortly describe the contribution in terms of content development performed by a group of history researchers from the third University of Rome.

\section{The Platform Architecture}

\begin{figure}
\begin{center}
\includegraphics[scale=0.5]{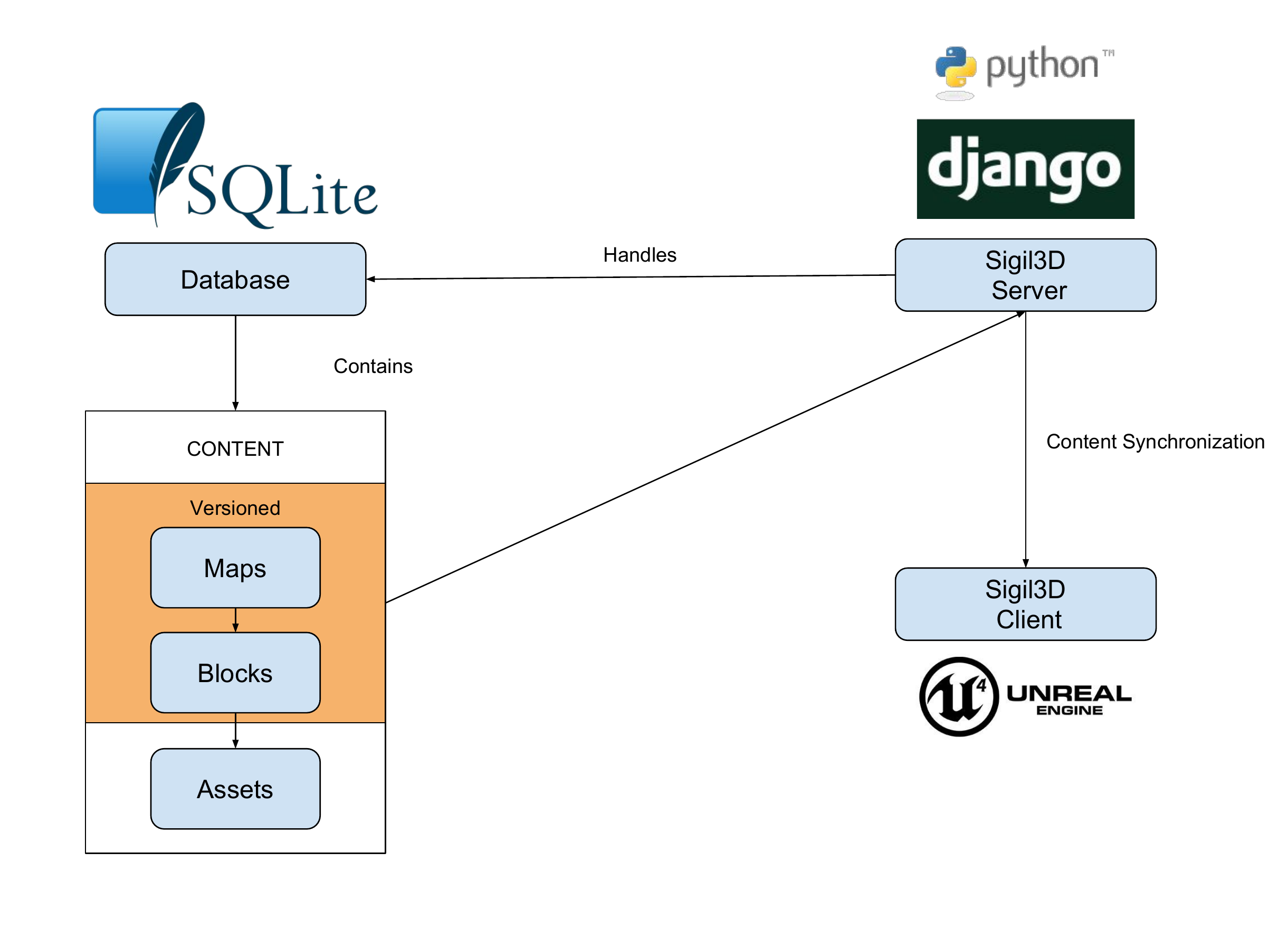}
\caption{Versioning system}
\label{fig:Versioning system}
\end{center}
\end{figure}

\begin{figure}
\begin{center}
\includegraphics[scale=0.5]{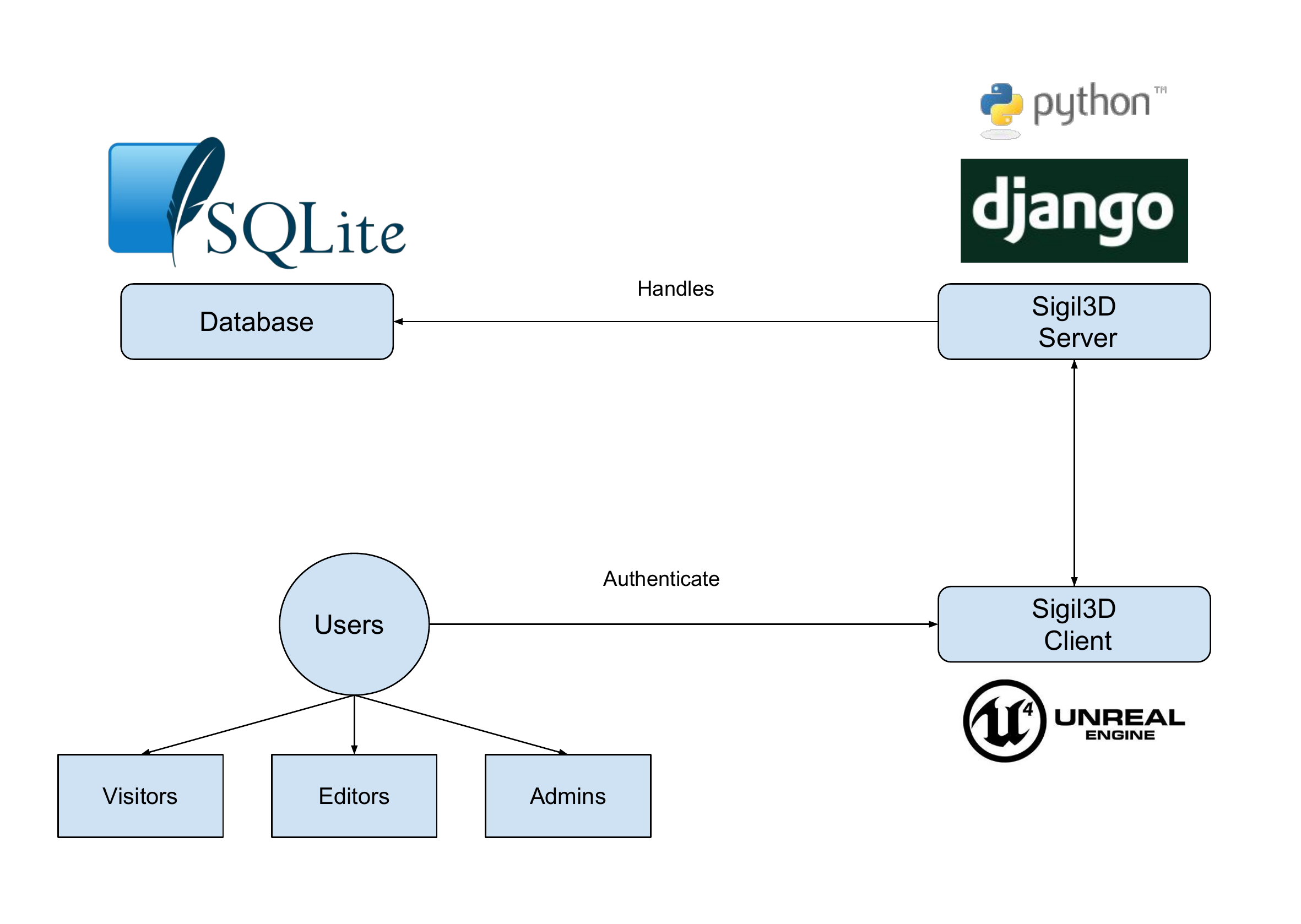}
\caption{Authentication}
\label{fig:Authentication}
\end{center}
\end{figure}

\begin{figure}
\begin{center}
\includegraphics[scale=0.5]{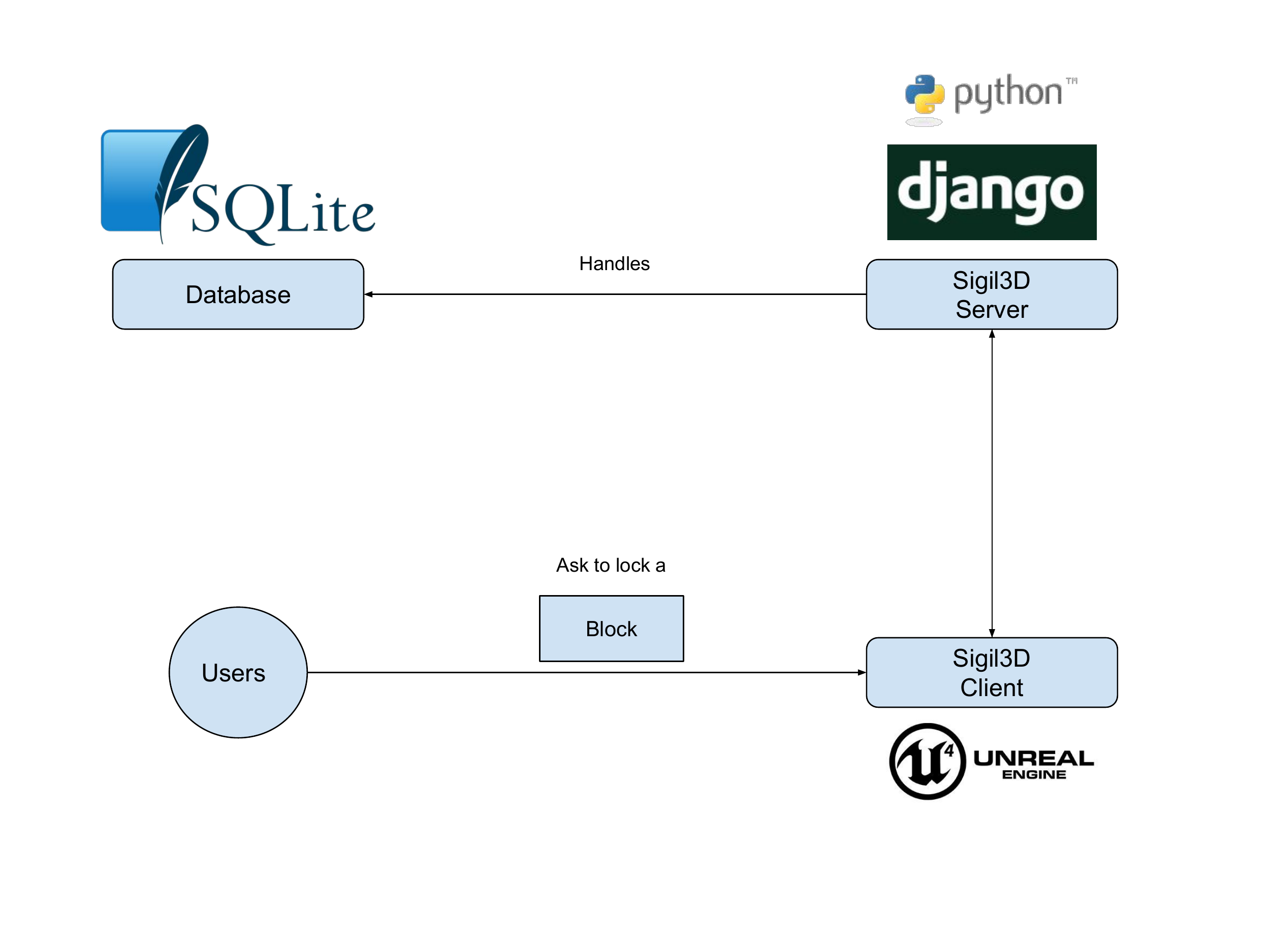}
\caption{Locking a block}
\label{fig:Locking a block}
\end{center}
\end{figure}

\begin{figure}
\begin{center}
\includegraphics[scale=0.5]{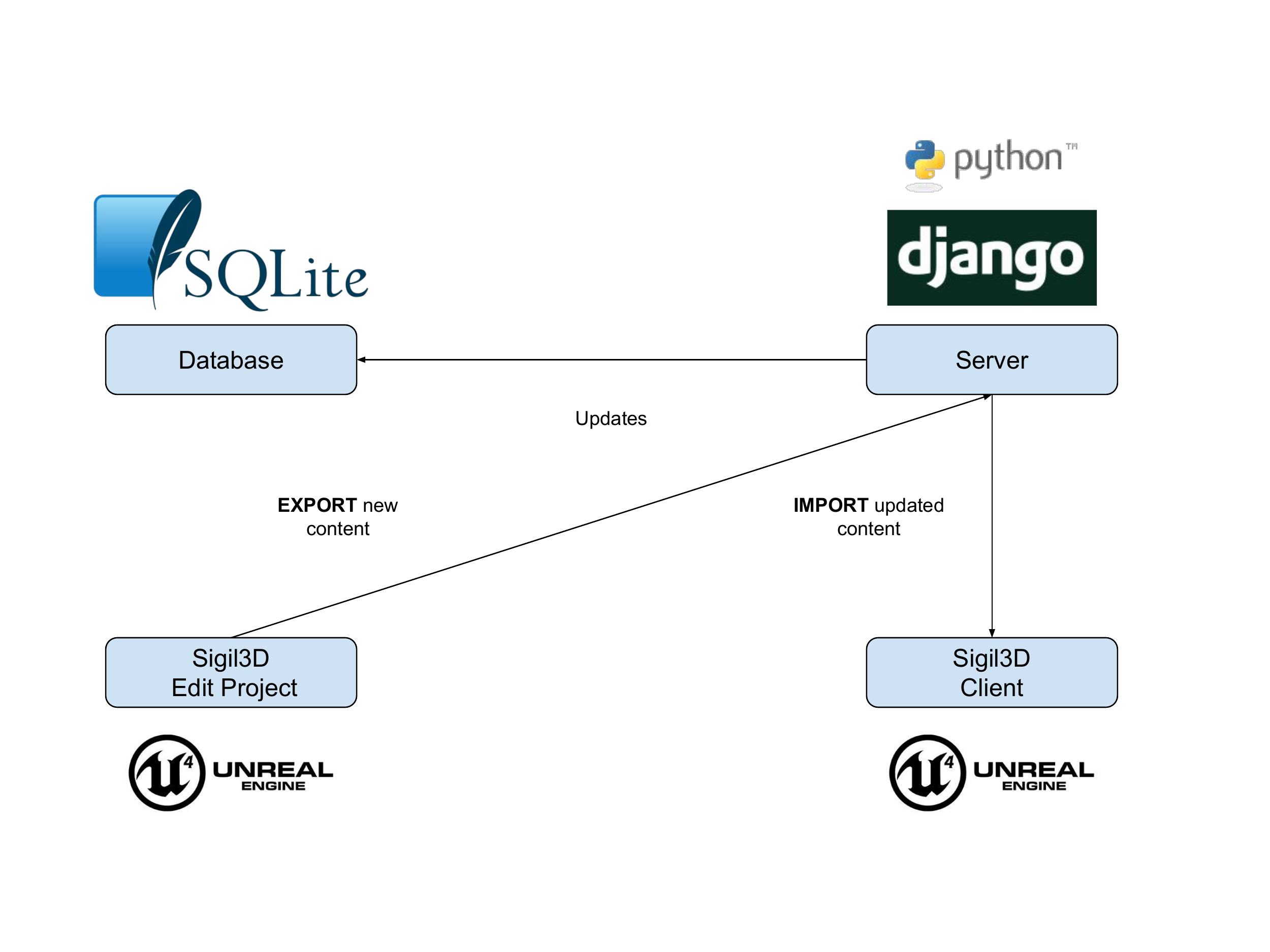}
\caption{Modifying an edit project}
\label{fig:Modifying an edit project}
\end{center}
\end{figure}

The platform architecture is built using four fundamental elements: a client, an edit project, a server, and a database (in addition to Unreal Engine 4 we also used the Django framework \cite{django}). The database contains all the content produced in the platform, which are stored as maps, and in turn contain blocks. The blocks which are loaded dynamically at runtime, are independent of the maps and can include Unreal assets that is resources of several types such as 3D models, textures, animations, and blueprints (i.e., visual script objects). A versioning system has been implemented (Figure~\ref{fig:Versioning system}), so that multiple versions can correspond to each map and block. The versioning system is controled by an administrator, who can approve or reject new versions. In this way, the user will always download and substitute the current local version with the latest version of a content. The users have to authenticate in order to modify the contents of the platform. This is done through an interaction with the Sigil3D server (Figure~\ref{fig:Authentication}). Currently, the platform includes three different roles: visitors (they can only view and interact with the maps), editors (they can modify the content of the blocks) and administrators (they build the maps, position the blocks, and approve the new versions). Once an editor is inside a map, they can ask to lock a specific block which is not already locked (Figure~\ref{fig:Locking a block}). Once the lock has been obtained, the editor is the only one who can modify the block. This operation is atomic in order to guarantee consistency between different modifications to the same content. When a specific block is locked, the editor can download an edit project, which will allow them to modify the content of the block ( Figure~\ref{fig:Modifying an edit project}). The edit project is a custom project from the Unreal Engine 4 editor, which intuitively acts like a plugin for this editor. Finally, once the Sigil3D edit project has been opened in the Unreal Engine 4 editor, the user will authenticate with the Sigil3D server in order to obtain the block whose content has to be modified. Once the modifications have been completed, the content of the Sigil3D edit project will be exported by executing an upload to the the Sigil3D server. The server controls the correctness of the new content and sends it to the database. When an administrator approves the modifications, the new content will be available to all visitors and editors. A demo of all this process is available in the following YouTube video: \texttt{https://www.youtube.com/watch?v=lpeq6PccYGY}.

\section{A Preliminary Experimental Evaluation}

In order to simulate a crowdsourcing interaction, we collaborated with Giuseppe Ragone and his research group at the third University of Rome, which fit perfectly with our objective of having someone who might be a future contributor to the platform, while avoiding 3D modeling specialists. The goal of this collaboration was to virtually reproduce a specific object from the \textit{Cyme} archaeological site, an ancient Aeolian city situated in modern Aliaga, Turkey. The specific object is an exedra, a semicircular recess created to let people sit and talk. It could be considered the ancient greek version of a bench. This exedra was placed inside an agora, which is the central space in the city-states of Ancient Greece. Today the only part of this exedra that remains intact is its base. In order to reconstruct a realistic and plausible version of the exedra as it was in its original state, the research group used information and pictures of other undamaged exedras in their possession. The group modeled the exedra static mesh by using the additive geometry tools provided by the Unreal Engine 4 editor. The group followed a couple of short tutorial videos to learn additive geometry and how to use the fundamental controls of the editor. They did not require other additional external help or assistance in order to produce the static mesh. To increment the static mesh realism we decided to add some texture to it. Unreal Engine 4 uses a collection of textures to generate material, which is an asset whose purpose is to paint the mesh. Since we did not want a generic stone material, the group asked for the help of a graphic designer to create a collection of textures that when putting them together would resemble to the actual stone of the \textit{Cyme} exedra. The final exedra static mesh was imported into the Sigil3D platform and can now be seen in the demo video previously cited. Overall, this experience shows that the Sigil3D usability is quite good. In particular, regarding the production and the editing of new content, the usability depends mostly on the usability of the Unreal Engine 4 editor, which in this case, turned out to be not so difficult for users that have no design or computer science skills.

\section{Future Developments}
We present several functions that can improve the platform usability. First, we should introduce an hub environment where admins may present their environments through virtual showcases which, also, should be used by visitors and editors to enter into these environments. Then, we should expand the role system so that admins can create independently an editor hierarchy inside their environments. This hierarchy should reflect onto block system, where an editor should add new smaller blocks inside a specific block and should assign permissions to edit these blocks to other users. They, in their turn, should manage their blocks independently. It should also be possible to build an asset repository as contributions grow. The platform could be used by editors to generate new, and more complex, 3D models. It could be used also by admins to provide templates to other admins so they can build new environments easily and quickly. Finally, the edit projects should be replaced by a standalone community tool through which users may create, modify and update environments, block and assets in a faster, smoother and more comfortable way.

The platform could be also integrated with virtual reality technologies, since Unreal Engine 4 supports several VR devices (such as Oculus Rift and HTC Vive) out of the box. Therefore Sigil3D virtual environments could be explored through this technology.

\end{document}